\documentclass[%
 reprint,
nofootinbib,
 amsmath,amssymb,
 aps,
]{revtex4-1}

\usepackage{graphicx}
\usepackage{dcolumn}
\usepackage{bm}

\begin{document}


\title{Gamma rays and neutrinos from a cosmic ray source in the Galactic Center region}

\author{A.~D.~Supanitsky}
\affiliation{Instituto de Astronom\'ia y F\'isica del Espacio (IAFE), CONICET-UBA, CC 67, Suc.~28, C1428ZAA Buenos Aires, Argentina.}
\email{supanitsky@iafe.uba.ar}

\date{\today}

\begin{abstract}

The center of the our Galaxy is a region where very energetic phenomena take place. In particular
powerful cosmic ray sources can be located in that region. The cosmic rays accelerated in these
sources may interact with ambient protons and/or low energy photons producing gamma rays and neutrinos.
The observation of these two types of secondary particles can be very useful for the identification
of the cosmic ray sources and for the understanding of the physical processes occurring during 
acceleration.

Motivated by the excess in the neutrino spectrum recently reported by the IceCube Collaboration, we study 
in detail the shape of the gamma ray and neutrino spectra originated from the interaction of cosmic ray protons 
with ambient protons for sources located in the Galactic Center region. We consider different models for proton 
acceleration and study the impact on the gamma ray and neutrino spectra. We also discuss the possibility to 
constrain and even identify a particular neutrino source by using the information given by the gamma ray spectrum 
taking advantage of the modification of the spectral shape, caused by the interaction of the gamma rays with the 
photons of the radiation field present in the interstellar medium, which strongly depends on the source distance.

\end{abstract}

\maketitle


\section{Introduction}

Gamma rays and neutrinos can be generated in cosmic ray sources due to the interaction of the cosmic 
rays with ambient photons and/or protons. Powerful cosmic ray accelerators can be located in the 
Galactic Center region; therefore, the observation of gamma rays and neutrinos coming from the Galactic 
Center is of great importance for the identification and understanding of the Galactic cosmic ray 
sources. It is worth noting that, in contrast with charged cosmic rays, these two types of secondary 
particles are not deviated by the magnetic fields present in the interstellar medium and then they 
point back to its source.   

The IceCube Collaboration has recently reported the observation of 28 neutrino events in the energy 
range from 30 TeV to 1.2 PeV \cite{IceCubeA:13,IceCubeB:13}, while only $10.6^{+4.6}_{-3.9}$ events
were expected considering a conventional atmospheric neutrino background. This corresponds to a 
$\sim 4.3\ \sigma$ excess. Despite the large angular uncertainty of these events ($\sim 10^\circ$),
we can observe that 5 of the 28 events come from the Galactic Center region. However, there is a 
degeneracy about the origin of these neutrinos because they can also originate outside our Galaxy.
It has been pointed out that the gamma ray flux associated with the neutrino flux can be useful to
constrain or even identify the candidate neutrino sources \cite{Gupta:12,Anchordoqui:13,Murase:13,Ahlers:13}.
In particular, in Refs.~\cite{Neronov:13,Razzaque:13} it is found that the the gamma ray data taken 
by Fermi LAT at lower energies are consistent with a Galactic origin of these five events. However,
the upper limits on the diffuse gamma ray flux obtained by different experiments at higher energies 
disfavor this hypothesis \cite{Ahlers:13}. 

On the other hand, the region in the cosmic ray energy spectrum where the transition from the Galactic 
to extragalactic cosmic rays takes place is still unknown (see Ref.~\cite{Aloisio:12} for a review).
Two possible regions for this transition are the second knee, a steepening of the spectrum which is 
given at an energy of $\sim 0.5$ EeV, and the ankle, a hardening of the spectrum placed at an energy of
$\sim 3$ EeV. In Ref.~\cite{Anchordoqui:13} it has been pointed out that the identification of Galactic 
neutrino sources and the observation of their spectra can contribute to finding the region where the 
transition from Galactic to extragalactic cosmic rays takes place. This is due to the fact that the end 
of the neutrino energy spectrum is correlated with the maximum energy at which the cosmic rays are 
accelerated in the source. Following Ref.~\cite{Anchordoqui:13}, the observation of the end of the neutrino 
spectrum, of a given Galactic source, at $E_\nu^{max} \cong 1$ PeV favors the scenario in which the transition 
takes place in the second knee region, whereas the observation of $E_\nu^{max} \cong 8-10$ PeV favors the 
scenario in which the transition is given at the ankle region.   
 
The most probable mechanism for the production of gamma rays and neutrinos in Galactic cosmic ray sources
is the interaction of cosmic rays with ambient protons \cite{Gupta:12}. In this work we study in detail
the shape of the gamma ray and neutrino spectra that originate from the proton-proton interaction in Galactic
cosmic ray sources placed in the Galactic Center region, which is motivated by the recent IceCube results.
We consider different shapes of the proton spectrum based on acceleration models. We also consider the 
values for the maximum energy of protons corresponding to the Galactic to extragalactic transition in the 
second knee region and in the ankle region. The propagation of the gamma rays and neutrinos from the Galactic 
Center to the Solar System is included in these calculations. While neutrinos suffer flavor oscillation during 
propagation, gamma rays can interact with the photons of the Galactic interstellar radiation field (ISRF) and 
the cosmic microwave background (CMB) which is the most important one \cite{Moskalenko:06}. In particular, we 
find that the ratio between the neutrino and gamma ray spectra depends on the cosmic ray proton spectrum, which 
can be used to study the conditions under which the cosmic ray protons are accelerated in the source. Finally, 
we study the possibility to discriminate between the Galactic and extragalactic origins of a neutrino spectrum 
observed in the direction of the Galactic Center by observing the associated gamma ray flux, which in principle 
will be possible with the planned high energy gamma ray observatories.

\section{Gamma rays and neutrinos injected by a cosmic ray source}
\label{Source}

Gamma rays and neutrinos can be produced as a result of the interaction of the cosmic rays 
with ambient protons present in the sources. Gamma rays are mainly produced by the decay 
of neutral pions; $\eta$ mesons also contribute to the gamma ray flux. The generation of
neutrinos is dominated by the decay of charged pions. Positive (negative) charged pions mainly 
decay into an antimuon and a muon (antimuon) neutrino,
\begin{eqnarray}
\pi^+ &\rightarrow& \mu^+ + \nu_{\mu},  \nonumber \\
\pi^- &\rightarrow& \mu^- + \bar{\nu}_{\mu}. \nonumber
\end{eqnarray}
The subsequent decay of the muons and antimuons produces more neutrinos,
\begin{eqnarray}
\mu^+ &\rightarrow& e^+ + \nu_{e} + \bar{\nu}_{\mu},  \nonumber \\
\mu^- &\rightarrow& e^- + \bar{\nu}_{e} + \nu_{\mu}. \nonumber
\end{eqnarray}

It is believed that the Galactic cosmic rays are accelerated principally at supernova remnants 
(see e.~g.~Ref.~\cite{Hillas:06}) by means of the first order Fermi mechanism. Also, the experimental 
evidence indicates that the injected composition of the Galactic sources is dominated by protons 
\cite{Longair}. The spectrum of protons that escape from the acceleration region accelerated via 
the first order Fermi mechanism, without taking account of energy losses, can be written as
\begin{eqnarray}
\Phi_p (E_p) &=& A\ E_p^{-\gamma}\ \left( 1+ \left( \frac{E_p}{E_{max}} \right)^{2 \delta} \right)  \nonumber \\
&& \times \exp\left( -\frac{\gamma-1}{2 \delta} \left( \frac{E_p}{E_{max}} \right)^{2 \delta} \right),
\label{Phip}
\end{eqnarray}
where $A$ is a normalization constant, $\gamma$ is the spectral index, $E_{max}$ refers to the
maximum energy which indicates the end of the spectrum (note that protons can be accelerated to higher 
energies than $E_{max}$), and $\delta$ comes from the energy dependence of the diffusion coefficient in 
the acceleration region, $k\propto E^\delta$ (see Appendix \ref{Acc} for details).  

Cosmic rays can suffer energy losses during the acceleration process due to the presence of strong
magnetic fields and ambient low energy photons \cite{Protheroe:98,Protheroe:04}. The spectrum near its
end can have different shapes, including pileups and sudden cutoffs, depending on the type of energy loss
dominating in the acceleration region. Then, in the subsequent analyses a proton spectrum with a sudden
cutoff in $E_{max}$ is also considered,
\begin{equation}
\label{V}
\Phi_p (E_p) = A \left\{ 
\begin{array}{ll}
  E_p^{-\gamma}  &  E_p\le E_{max} \\
                 &               \\
  0              &  E_p>E_{max} 
\end{array}  \right..
\end{equation}

\begin{figure}[!ht]
\includegraphics[width=8cm]{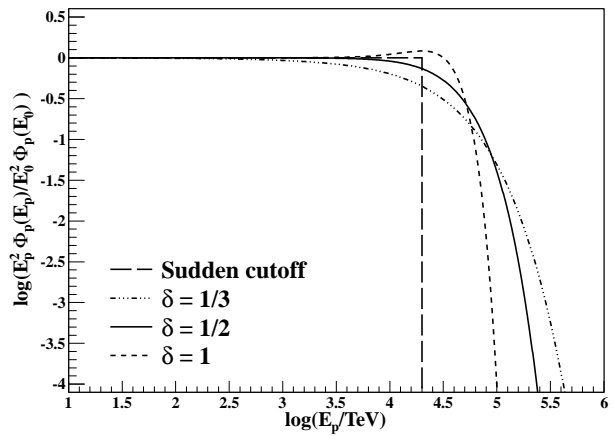}
\caption{\label{ProtonS} Logarithm of the proton spectra multiplied by $E^2$ as a function of the logarithm 
of the proton energy. The spectral index considered is $\gamma = 2$ and the maximum energy is 
$E_{max}=2\times10^4$ TeV. The spectra multiplied by $E^2$ are normalized at an energy $E_{0} = 10$ TeV.}
\end{figure}

Figure \ref{ProtonS} shows the proton spectra for $\gamma=2$ and for $\delta = 1/3$ (Kolmogorov spectrum), 
$\delta = 1/2$ (Kraichnan spectrum), and $\delta = 1$ (completely disordered field) \cite{Protheroe:04}. 
The spectrum corresponding to the sudden cutoff is also shown. The maximum energy considered is 
$E_{max} = 2\times10^4$ TeV, which corresponds to the transition between the Galactic and extragalactic 
cosmic rays in the region of the second knee. As expected, the steepening of the spectra is more pronounced 
for increasing values of $\delta$.

Given the proton spectrum the spectra of gamma rays and neutrinos generated by the interaction with the ambient 
protons are calculated following Ref.~\cite{Kelner:06}. In that work the proton-proton interaction was simulated
by using the hadronic interaction model Sibyll 2.1 \cite{Sibyll}. The gamma ray spectrum is calculated from
\begin{equation}
\Phi_\gamma^0 (E_\gamma) = C \int_{E_\gamma}^\infty \frac{dE_p}{E_p}\ \sigma_{inel}(E_p)\ \Phi_p (E_p)\ %
F_{\gamma}\left( \frac{E_\gamma}{E_p},E_p \right), 
\label{Fgamma}
\end{equation}
where $C$ is a normalization constant and $\sigma_{inel}$ is the proton-proton inelastic cross section. $\sigma_{inel}$ 
and $F_{\gamma}$ are taken from Ref.~\cite{Kelner:06}. The neutrino and antineutrino spectra are obtained in a similar 
way, and the corresponding functions are also taken from Ref.~\cite{Kelner:06}.

\begin{figure}[!ht]
\includegraphics[width=8cm]{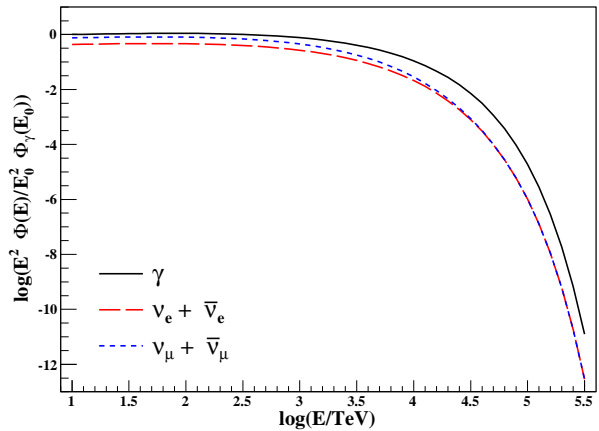}
\caption{Logarithm of the gamma ray and neutrino spectra multiplied by $E^2$ as a function of the logarithm           
of the energy for $\gamma = 2$, $E_{max}=2\times10^4$ TeV, and $\delta=1/2$. Also in this case $E_{0} = 10$ TeV. 
\label{GNSource}}
\end{figure}

Figure \ref{GNSource} shows the gamma ray and neutrino\footnote{Hereafter, neutrino spectrum will refer to the sum of the 
neutrino and antineutrino spectra of a given flavor.} spectra at the source for $\gamma=2$, $E_{max} = 2\times10^4$ TeV, 
and $\delta =1/2$. Note that the results obtained in this work are in very good agreement with the ones obtained in 
Ref.~\cite{Kappes:07}.
From Fig.~\ref{GNSource} it can be seen that the gamma ray spectrum is larger than the one corresponding to the electron 
neutrino and muon neutrino. Also, the end of the neutrino spectra takes place at a smaller energy than the one corresponding  
to gamma rays.  

\section{Gamma rays and neutrinos at Earth}

\subsection{Precise calculation}
\label{NC}

Gamma rays and neutrinos produced at a given source propagate through the interstellar medium to reach the Earth.
The electron and muon neutrinos oscillate during propagation; therefore, the flavor ratio observed at the Earth is different
from the one in the source. It is believed that the neutrino oscillations originate from the fact that the flavor states,
$|\nu_\alpha \rangle$ with $\alpha=e,\ \mu,\ \tau$, are not the mass eigenstates, $|\nu_j \rangle$ with $j=1,\ 2,\ 3$.
The flavor and the mass basis are related by \cite{Giunti:07}
\begin{equation}
\label{Mix}
|\nu_\alpha \rangle = \sum_{j=1}^3 U^*_{\alpha j}|\nu_j \rangle,
\end{equation}
where $U$ is a unitary matrix which can be parametrized by the mixing angles $\theta_{12}, \theta_{23},$ and $\theta_{13}$ 
and a $CP$ violating phase $\delta_{CP}$. The transition probability can be written as \cite{Giunti:07}
\begin{equation}
\label{TP}
P_{\nu_\alpha \rightarrow \nu_\beta} = \sum_{j=1}^3 |U_{\alpha j}|^2\ |U_{\beta j}|^2.
\end{equation}
In the so-called tribimaximal mixing approximation \cite{Harrison:02} the parameters of the mixing matrix are taken as 
$\theta_{12}=\arcsin(1/\sqrt{3})$, $\theta_{23}=\pi/4$, $\theta_{13}=0$, and $\delta_{CP}=0$. By using this approximation 
it can be seen that a flavor ratio $1:2:0$ in the source goes to $1:1:1$ after propagation. In this work a more precise 
estimation of the mixing angles is used in which $\theta_{13}$ is different from zero \cite{PDG:12}: 
$\sin^2(2\theta_{12}) = 0.857$, $\theta_{23}=\pi/4$, and $\sin^2(2\theta_{13})=0.096$. It is also assumed that $\delta_{CP}=0$.  
By using this new estimation of the parameters the flavor ratio $1:2:0$ goes to $1.05:0.99:0.95$, which is very close to the
tribimaximal mixing approximation. Therefore, the propagated spectra are calculated from
\begin{equation}
\label{Spec}
\left(\begin{array}{c}
  \Phi_{\nu_e}    \\
  \Phi_{\nu_\mu}  \\
  \Phi_{\nu_\tau} \\
\end{array}\right) =
\bm{P}%
\left(\begin{array}{c}
  \Phi_{\nu_e}^0    \\
  \Phi_{\nu_\mu}^0  \\
  \Phi_{\nu_\tau}^0 \\
\end{array}\right)
\end{equation}
where the elements of the $\bm{P}$ matrix are given by $\bm{P}_{\alpha\beta}=P_{\nu_\alpha \rightarrow \nu_\beta}$ and 
$\Phi_{\nu_\alpha}^0$ are the neutrino spectra at the source. Note that in this case $\Phi_{\nu_\tau}^0=0$.

\begin{figure}[!ht]
\includegraphics[width=8cm]{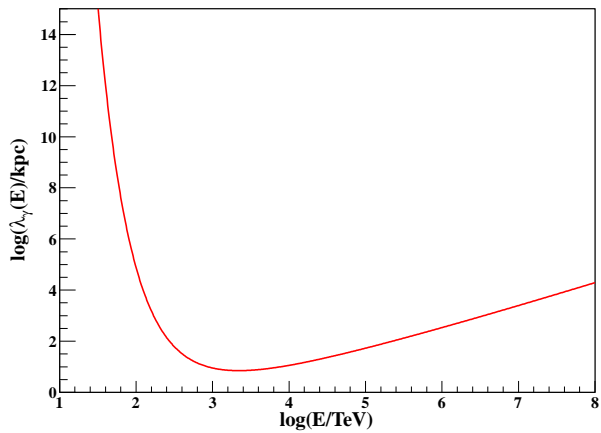}
\caption{Logarithm of the mean free path of gamma rays as a function of the logarithm of the primary energy for electron-positron pair 
production in the CMB. \label{Lcmb}}
\end{figure}

Gamma rays interact with photons of different backgrounds when they propagate from the source to the Earth; the most important 
ones are the CMB and the ISRF \cite{Moskalenko:06}. The most important process at the energies considered is the electron-positron 
pair production. Figure \ref{Lcmb} shows the mean free path of photons, $\lambda_\gamma$, as a function of energy for the CMB. 
The calculation has been done following Ref.~\cite{Protheroe:86}. From the figure it can be seen that for energies of order of 1 PeV 
$\lambda_\gamma\sim 9$ kpc, which is very close to the distance from the Solar System to the Galactic Center ($D=8.5$ kpc). 
Therefore, at these energies the attenuation in the CMB becomes quite important.   

\begin{figure}[!ht]
\includegraphics[width=8cm]{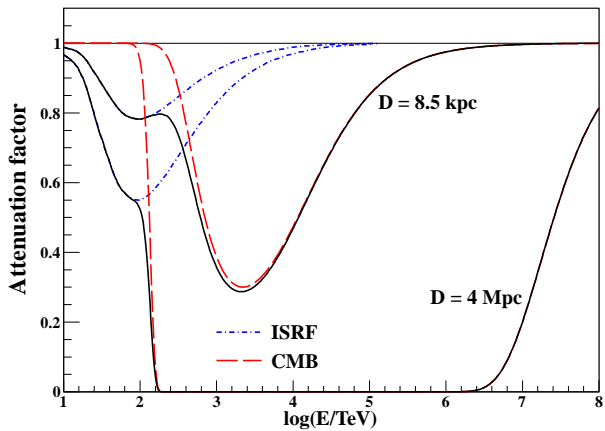}
\caption{Attenuation factor of gamma rays as a function of the logarithm of energy. The photon backgrounds considered are the CMB and the 
ISRF. One source is placed at the Galactic Center ($D=8.5$ kpc), and the other is placed at $D=4$ Mpc in the direction of the Galactic 
Center. 
\label{AFact}}
\end{figure}

The attenuation factor is given by $T(E_\gamma,D) = \exp(-\tau(E_\gamma,D))$, where $D$ is the distance from the source to 
the Earth and
\begin{equation}
\label{AttFact}
\tau(E_\gamma,D) = \int_0^D \frac{d\ell}{\lambda_\gamma(E_\gamma,\ell)}.
\end{equation}
Thus, the gamma ray flux at the Earth is given by
\begin{equation}
\label{AttFact2}
\Phi_\gamma (E_\gamma) = \frac{T(E_\gamma,D)\ \Phi_\gamma^0 (E_\gamma)}{4 \pi D^2}.
\end{equation}
Figure \ref{AFact} shows the attenuation factor for the CMB and ISRF (taken from Ref.~\cite{Moskalenko:06}) as a function of the energy 
for two different positions of the source: One is placed at the Galactic Center and the other at 4 Mpc from the Solar System in the 
direction of the Galactic Center. The attenuation factor for the ISRF is taken from Ref.~\cite{Moskalenko:06}; the one corresponding 
to the extragalactic source is obtained by integrating $\lambda_{\gamma}^{-1}( E_{\gamma} , \ell )$ from a distance of 28.5 kpc in the 
direction of the Galactic Center (see Eq.~(\ref{AttFact})). It can be seen that the attenuation caused by the interaction with the CMB 
photons is more important than that corresponding to the ISRF. 

\begin{figure}[!ht]
\includegraphics[width=8cm]{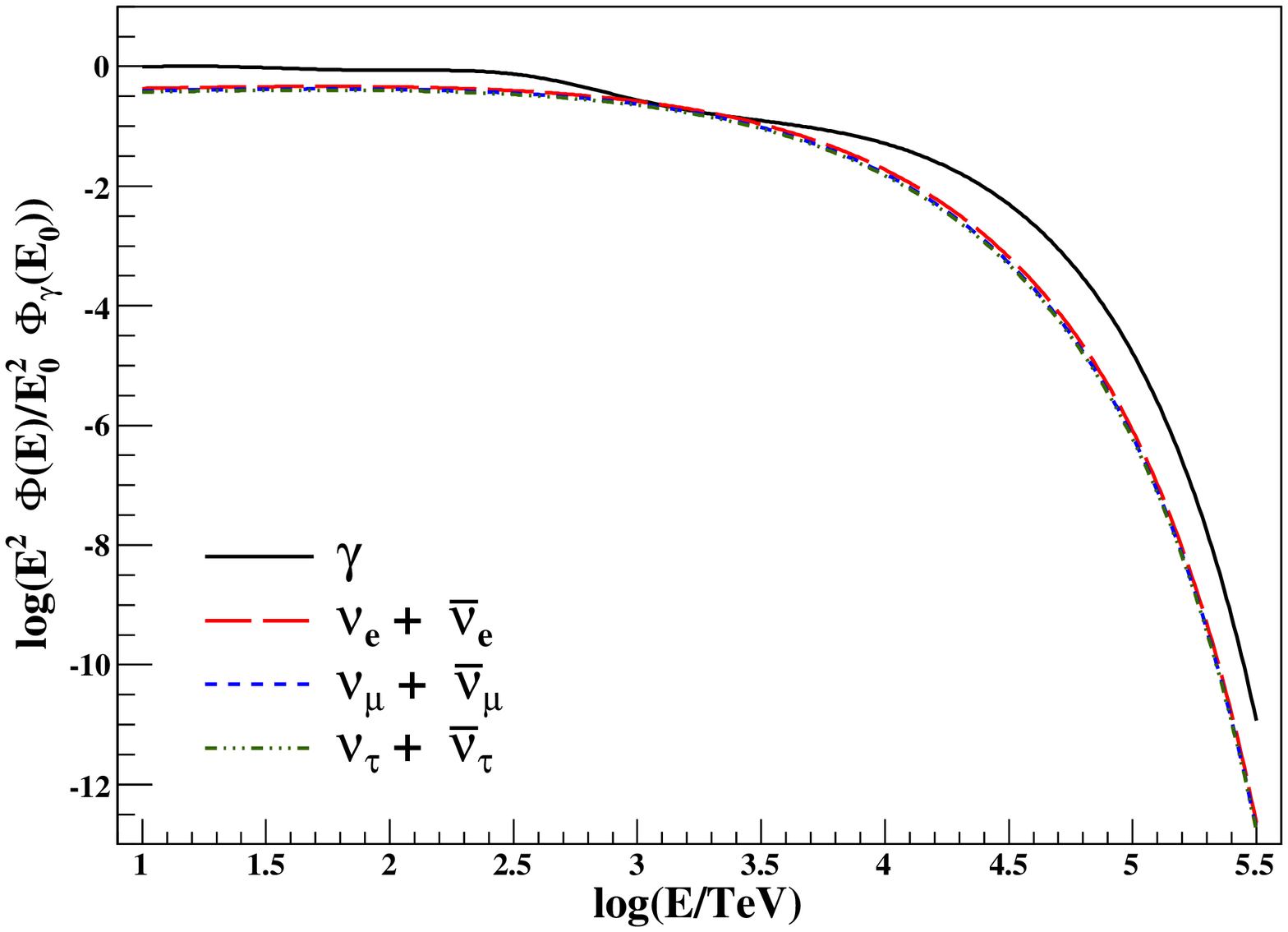}
\includegraphics[width=8cm]{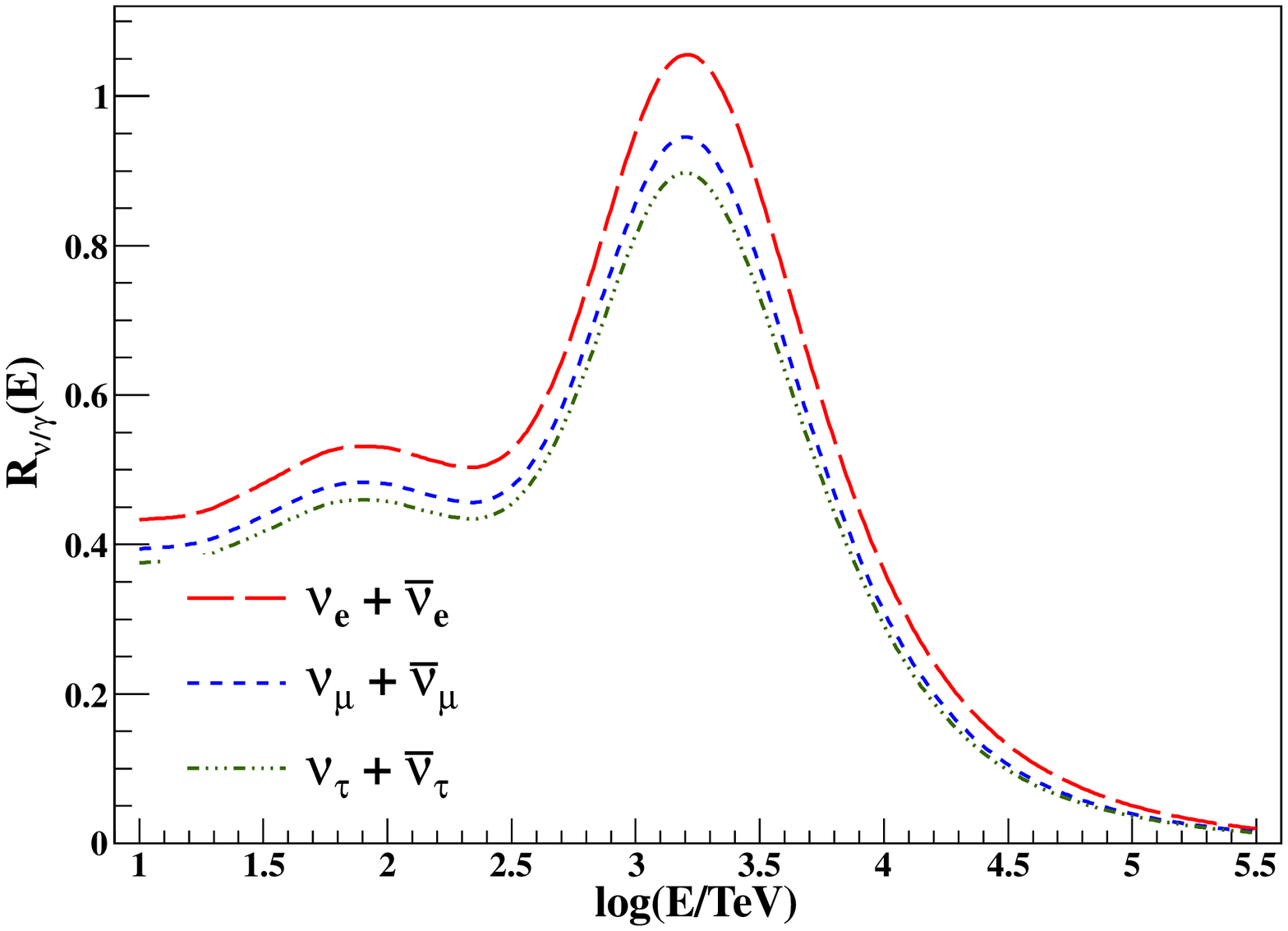}
\caption{Top panel: Logarithm of the gamma ray and neutrino spectra at the Earth multiplied by $E^2$ as a function of the logarithm
of the energy for a cosmic ray source located at the Galactic Center. Bottom panel: Ratio of the neutrino spectra to the gamma ray 
spectrum as a function of the logarithm of the energy. The parameters used are $\gamma = 2$, $E_{max}=2\times10^4$ TeV, and 
$\delta =1/2$.\label{PropSpec}}
\end{figure}

The top panel of Fig.~\ref{PropSpec} shows the gamma ray and neutrino spectra at the Earth for a Galactic cosmic ray source, like the one 
considered in Sect.~\ref{Source} ($\gamma=2$, $E_{max} = 2\times10^4$ TeV, and $\delta =1/2$), located in the Galactic Center. In this 
figure the effects of the attenuation in the gamma ray spectra are evident. Also, the neutrino spectra for the different flavors become 
similar after propagation. The bottom panel of Fig.~\ref{PropSpec} shows the ratio $R_{\nu/\gamma}$ of the neutrino spectra to the 
gamma ray spectrum. It can be seen that it is not constant in a large energy range due to the attenuation suffered by the gamma rays and 
the different ends of the neutrino and gamma ray spectra. In particular, the first maximum of $R_{\nu/\gamma}$ corresponds to the
attenuation of the gamma rays in the ISRF, whereas the second and larger one corresponds to the attenuation in the CMB. 

\begin{figure}[!ht]
\includegraphics[width=8cm]{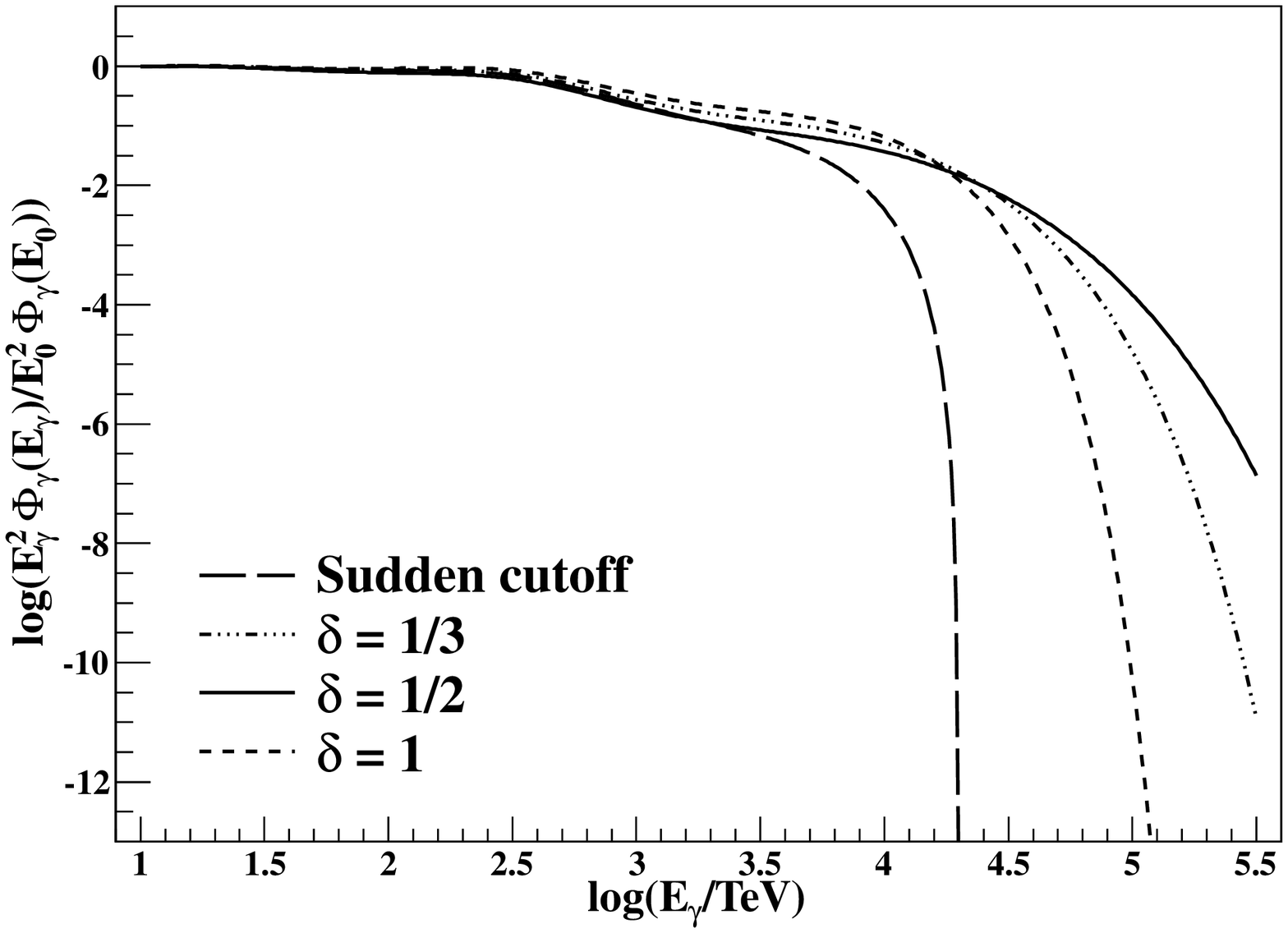}
\includegraphics[width=8cm]{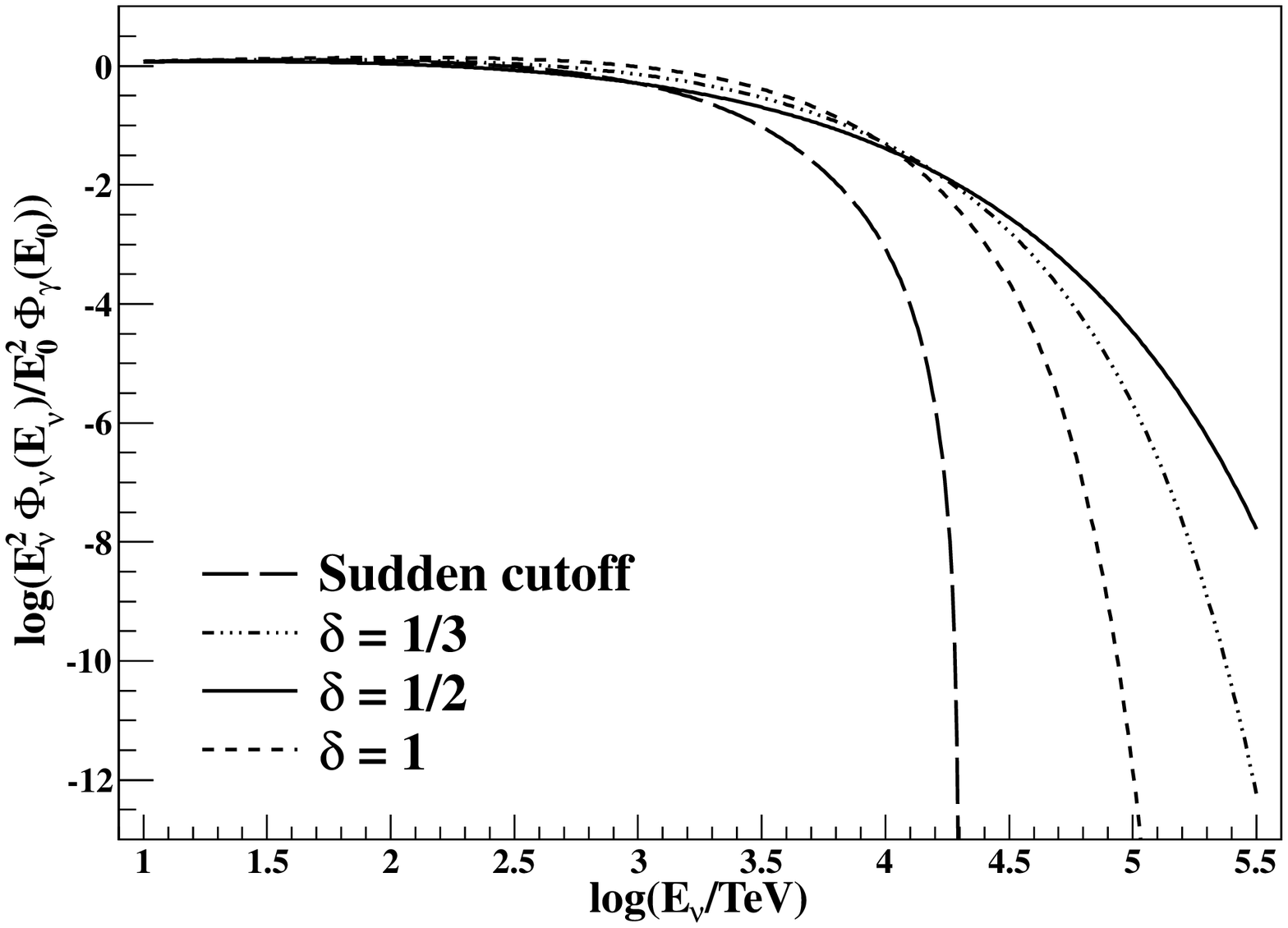}
\caption{Logarithm of the gamma ray (top panel) and neutrino (bottom panel) spectra at the Earth multiplied by $E^2$ as a function of the 
logarithm of the energy for a cosmic ray source located at the Galactic Center and for the different cases of the proton spectrum
considered. The maximum energy of the proton spectrum is $E_{max}=2\times10^4$ TeV.  \label{SpecEnd}}
\end{figure}

The shape of the end of the gamma ray and neutrino spectra depends on the shape of the proton spectrum. Figure \ref{SpecEnd} shows 
the gamma ray and the total neutrino spectra (summed over all flavors) at the Earth as a function of energy for the different values of 
$\delta$ considered and for the case in which the proton spectrum presents a sudden cutoff. The maximum energy of the proton spectrum 
is $E_{max}=2\times10^4$ TeV. As expected, when the proton spectrum ends faster, the gamma ray and the neutrino spectra also end faster.   

\begin{figure}[!ht]
\includegraphics[width=8cm]{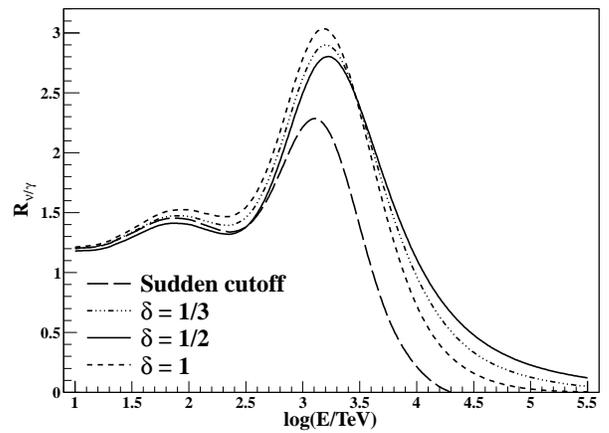}
\caption{Ratio between the neutrino spectrum and the gamma ray spectrum as a function of the logarithm of the energy for a cosmic ray source
located in the Galactic Center region for the different cases of the proton spectrum considered. The maximum energy of the proton spectrum 
is $E_{max}=2\times10^4$ TeV. \label{RatioD}}
\end{figure}

Figure \ref{RatioD} shows the ratio between the total neutrino spectrum and the gamma ray spectrum as a function of energy, corresponding 
to the spectra shown in Fig.~\ref{SpecEnd}. It can be seen that $R_{\nu/\gamma}$ presents two maxima; as mentioned above, the first one 
corresponds to the attenuation due to the interaction of the gamma rays with the photons of the ISRF and the second one to the CMB. Also, 
the shape of $R_{\nu/\gamma}$ depends on the type of proton spectrum considered, which can be useful, from the observational point of view, 
to study the end of the proton spectrum at the source, which is intimately related to the diffusion coefficient and the energy losses
suffered by protons in the acceleration region.    

\begin{figure}[!ht]
\includegraphics[width=8cm]{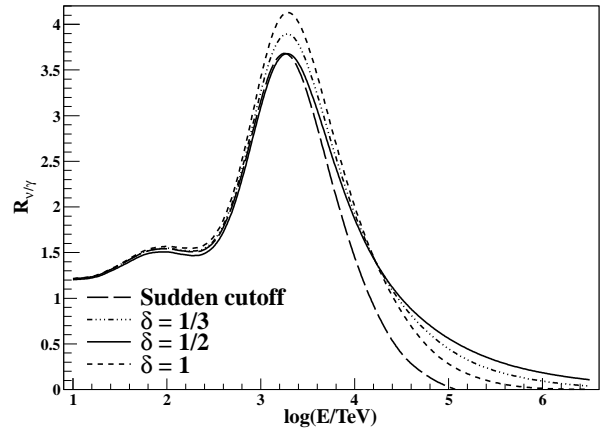}
\caption{Ratio between the neutrino spectrum and the gamma ray spectrum as a function of the logarithm of the energy for a cosmic ray source
located in the Galactic Center region for the different cases of the proton spectrum considered. The maximum energy of the proton spectrum 
is $E_{max}=1.2\times10^5$ TeV. \label{RatioDAnk}}
\end{figure}

The gamma ray and neutrino spectra are modified when the maximum energy of the accelerated protons increases. Figure \ref{RatioDAnk}
shows the ratio between the total neutrino spectrum and the gamma ray spectrum as a function of energy for the maximum energy of the 
proton spectrum $E_{max}=1.2\times10^5$ TeV, which corresponds to the case in which the transition between the Galactic and extragalactic
cosmic rays is given in the ankle region. It can be seen that in this case $R_{\nu/\gamma}$ is larger than one corresponding to the case 
considered above. This is due to the fact that, in this case, the region in which the attenuation on the CMB is more important 
($E_\gamma \sim 10^{3.3}$ TeV), the gamma ray and neutrino spectra approximately follow the shape of the proton spectrum; i.e.~the end 
of the spectra is given at higher energies. Also in this case, the ratio can be useful to study the end of the proton spectrum at the 
source, which is related to the physical conditions under which the protons are accelerated.  

It can be seen that by increasing the spectral index $\gamma$, the steepening of the gamma ray and neutrino spectra takes place at lower 
energies, which makes $R_{\nu/\gamma}$ decrease.

\subsection{Approximate calculation}

It is quite common to calculate the gamma ray spectrum from the neutrino spectrum (and the other way around), making use of several 
approximations (see, for instance, Ref.~\cite{Ahlers:13}). In this section the accuracy of this approximation is studied by means of 
the numerical calculations presented above. In proton-proton collisions $\pi^0$, $\pi^+$, and $\pi^-$ are created with 
the same multiplicity. As a result of the decay of charged pions and the subsequent muon or antimuon decay four particles are 
created; three of those are neutrinos, and then $E_\nu \cong E_\pi/4$. In the case of neutral pion decay two photons are created; then 
$E_\gamma \cong E_\pi/2$. Therefore, $E_\nu \cong E_\gamma/2$, and the differential flux of gamma rays can be obtained from 
\cite{Ahlers:13}
\begin{eqnarray}
\Phi_\gamma(E_\gamma) &\cong& \frac{1}{3} \sum_\alpha \Phi_{\nu_\alpha+\bar{\nu}_\alpha} \!\! \left(\frac{E_\gamma}{2} \right)\ %
\frac{d E_\nu}{d E_\gamma} \nonumber \\
&\cong& \frac{1}{6} \sum_\alpha \Phi_{\nu_\alpha+\bar{\nu}_\alpha} \!\! \left(\frac{E_\gamma}{2} \right),
\label{PhiApp}
\end{eqnarray}   
where $\alpha = e, \mu, \tau$. Here the effects of the propagation of gamma rays in the interstellar medium are not taken into account
in order to better compare the spectral shape of the numerical calculation and the approximation. 

\begin{figure}[!ht]
\includegraphics[width=8cm]{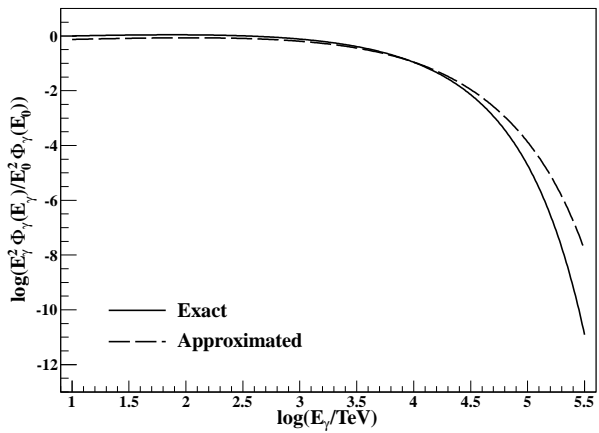}
\caption{Logarithm of the gamma ray spectra multiplied by $E^2$ as a function of the logarithm of the energy for $\gamma = 2$,
$E_{max}=2\times10^4$ TeV, and $\delta =1/2$. The solid line corresponds to the numerical calculation and the dashed line to the 
approximation based on Eq.~(\ref{PhiApp}). \label{FGammaApp}}
\end{figure}

Figure \ref{FGammaApp} shows the gamma ray spectrum as a function of the energy calculated as in Sec.~\ref{NC} (solid line) and from 
the neutrino spectra, also calculated as in Sec.~\ref{NC}, by using Eq.~(\ref{PhiApp}) (dashed line) for $\gamma = 2$, 
$E_{max}=2\times10^4$ TeV, and $\delta =1/2$. The approximation underestimates the gamma ray spectrum at lower energies and overestimates 
the spectrum at higher energies. Note that the larger differences between the approximated calculation and the numerical one are given at
high energies, in the region at the end of the spectrum. In particular, the approximated spectrum is less steep in that energy region,
which makes the difference between the approximated spectrum and the numerical one to increase with energy.    

\begin{figure}[!ht]
\includegraphics[width=8cm]{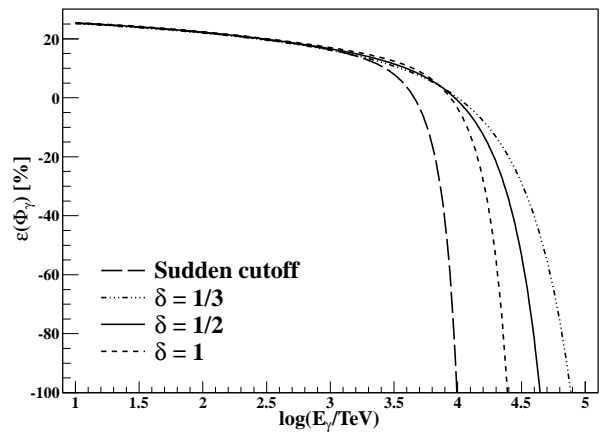}
\caption{Relative error of the approximated spectra as a function of the logarithm of the energy for $\gamma = 2$ and 
$E_{max}=2\times10^4$ TeV.\label{DeltaGammaApp}}
\end{figure}

In order to quantify these differences let us introduce the following parameter,
\begin{equation}
\varepsilon(\Phi_\gamma)(E) = 1-\frac{\Phi_\gamma^{app}(E)}{\Phi_\gamma(E)},
\end{equation}
where $\Phi_\gamma^{app}(E)$ is the approximated spectrum and $\Phi_\gamma(E)$ is the one obtained numerically. Figure 
\ref{DeltaGammaApp} shows the relative error $\varepsilon(\Phi_\gamma)$ as a function of energy for the four cases of the proton
spectrum considered with $\gamma = 2$ and $E_{max}=2\times10^4$ TeV. At lower energies the differences between the approximated spectra 
and the numerical one are smaller than $\sim 26$ \%. The larger differences take place in the region at the end of the gamma ray spectra
because, as mentioned above, the approximated spectra is less steep than the numerical one. Note that in all cases considered, the behavior
of the relative error is qualitatively similar; in the region corresponding to the end of the spectra, the absolute value of the relative 
error increases with energy, reaching values of the order of $100\%$ very fast. Note also that the approximation underestimates the spectrum 
at lower energies and overestimates it in the region corresponding to the end of the spectrum, i.e.~$\varepsilon(\Phi_\gamma)$ changes its 
sign at a given energy depending on the proton spectrum considered.  

\section{Galactic versus extragalactic origin of the observed neutrino spectrum}

It is not possible to elucidate the origin of neutrinos coming from the Galactic Center region with just the observation of
their energy spectrum and arrival direction distribution. There is a degeneracy; these neutrinos can originate in other regions
of the Galaxy or even outside the Galaxy. This degeneracy can be broken by observing the companion gamma ray flux because the 
attenuation of the gamma rays on the CMB and ISRF depends on the distance of the source.  

\begin{figure}[tb]
\includegraphics[width=8cm]{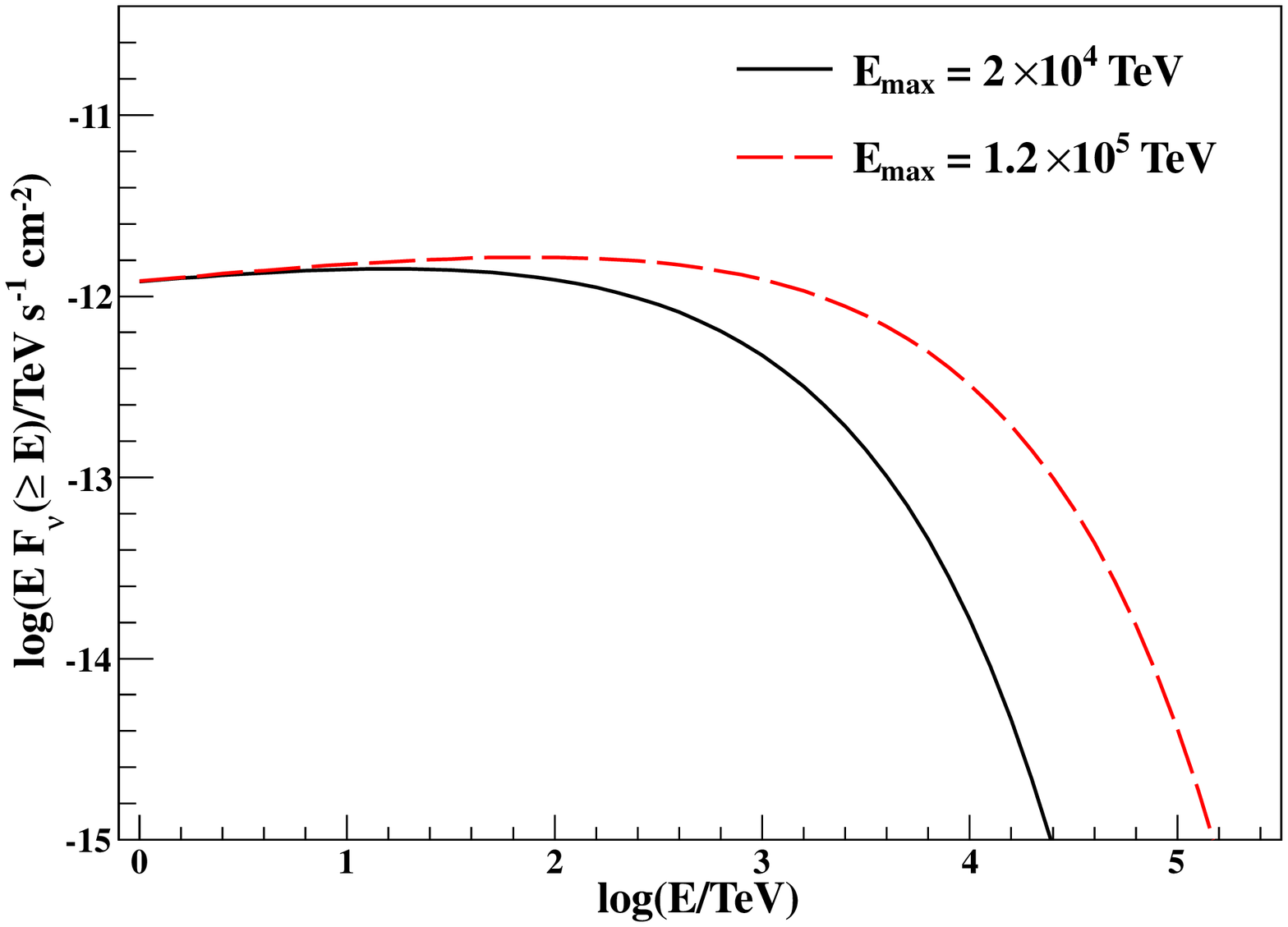}
\includegraphics[width=8cm]{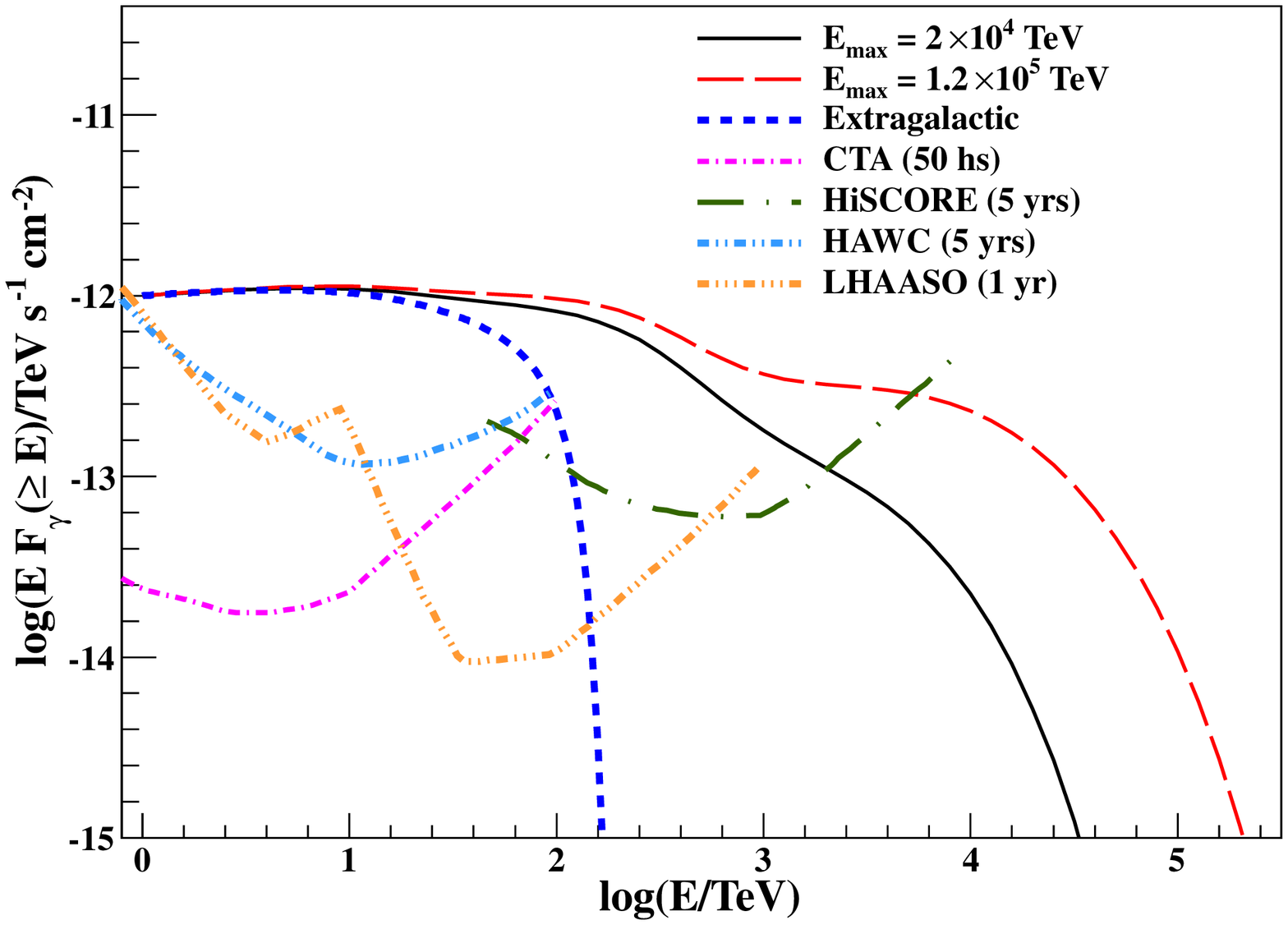}
\caption{Top panel: Logarithm of the integral neutrino flux multiplied by the energy as a function of the logarithm of the energy for
two values of the maximum energy of the proton spectrum. Bottom panel: Logarithm of the integral gamma ray flux multiplied by the energy 
as a function of the logarithm of the energy for a source located in the Galactic Center for two different values of the maximum energy 
of the proton spectrum and for an extragalactic source located at 4 Mpc in the direction of the Galactic Center. Also shown are the 
sensitivity curves of different planned or under-construction experiments.
\label{IntFlux}}
\end{figure}

The top panel of Fig.~\ref{IntFlux} shows the integral spectra at the Earth of neutrinos (all flavors) produced in proton-proton 
interactions for the two values of the maximum energy of the proton spectrum considered, for $\gamma = 2$, and $\delta=1/2$. The 
bottom panel of Fig.~\ref{IntFlux} shows the corresponding gamma ray flux for these two cases and also for the case in which the 
gamma rays and neutrinos originate in an extragalactic source. For the extragalactic case it is assumed that the source is located 
at 4 Mpc from the Solar System; the attenuation in the CMB and ISRF used for the calculation corresponds to the extragalactic case of 
Fig.~\ref{AFact}. Also in this case, we use $\delta=1/2$. The maximum energy for the proton spectrum used in this case 
is $E_{max}=2\times 10^4$ TeV, but a very similar spectrum is obtained for $E_{max}=1.2\times 10^5$ TeV due to the sudden cutoff at 
$\sim 100$ TeV produced by the CMB and ISRF attenuation. The normalization of the spectra is chosen such that the gamma ray flux is 
very well observed by the different experiments. The bottom panel of Fig.~\ref{IntFlux} also shows the sensitivity of different 
planned or under-construction experiments: CTA (Cherenkov Telescope Array) \cite{cta}, HiSCORE (Hundred*i Square-km Cosmic ORigin 
Explorer) \cite{hiscore}, HAWC (High Altitude Water Cherenkov) \cite{hawc}, and LHAASO (Large High Altitude Air Shower Observatory) 
\cite{lhaaso}. Note that only CTA and HiSCORE are planned for the Southern Hemisphere\footnote{The CTA project consists in the 
construction of two observatories, one in each hemisphere, but the southern observatory will be constructed first.} which is ideal 
for the observation of the Galactic Center.  

From Fig.~\ref{IntFlux} it can be seen that, thanks to the attenuation of the gamma rays in the photon fields, it is possible to 
at least decide about the Galactic or extragalactic origin of the neutrino flux observed at the Earth. Moreover, in the case of a 
Galactic neutrino source not necessarily placed in the Galactic Center region, the measurement of the corresponding gamma ray flux 
can give an estimation of the distance of the source, provided that the gamma ray flux reaches energies larger than a few PeV.  

It is worth noting that the test proposed here requires clear evidence of the existence of a neutrino source and a good 
measurement of its spectrum. Although the present IceCube data show some hints of the existence of one or more neutrino sources 
in the Galactic Center region, the present statistics is still poor for this type of test. This situation can change in the near
future when more data are collected.

\section{Conclusions}

In this work we have studied in detail the shape of the gamma ray and neutrino energy spectra generated by the interaction of
cosmic ray protons with ambient protons. We have considered two possible maximum energies for the proton spectrum: one consistent 
with the Galactic to extragalactic transition of the cosmic rays given in the second knee region and the other in the ankle 
region. We have studied the propagation of gamma rays and neutrinos from the source to the Earth, paying special attention to the 
case in which the source is located in the Galactic Center region. We have shown that the shape of the gamma ray and neutrino spectra 
at the Earth reflects the properties of the spectrum of protons accelerated in the sources; then the observation of these two types of 
secondaries can be very useful for the study of the cosmic ray sources. 

We have studied the accuracy of a common approximation to infer the gamma ray spectrum from the neutrino spectra. We have found 
that at low energies the approximate calculation underestimates the gamma ray spectrum in less than $\sim 26$ \% (for the four different 
types of proton spectra considered in this work) and that at higher energies the approximate calculation overestimates the gamma ray 
spectrum. The differences between the approximate spectrum and the one obtained numerically are larger in the region corresponding to 
the end of the spectrum and increase with energy. This is due to the fact that the approximate spectrum is less steep than the one 
obtained numerically. 

We have also discussed the existing degeneracy about the origin of neutrinos observed in the direction of Galactic Center region. We 
have shown that such degeneracy can be broken by observing the companion gamma ray spectrum (above a few PeV), which strongly depends on 
the distance to the source due to the interaction of the gamma rays with the low energy photons of the CMB and ISRF.

\begin{acknowledgments}

A.~D.~S.~is a  member of the Carrera del Investigador Cient\'ifico of CONICET, Argentina. This work is supported by CONICET PIP 
114-201101-00360 and ANPCyT PICT-2011-2223, Argentina. 

\end{acknowledgments}

\appendix

\section{Energy spectrum of cosmic ray protons}
\label{Acc}

It is believed that cosmic rays are accelerated by the diffusive shock acceleration mechanism. Following 
Refs.~\cite{Protheroe:98,Drury:99,Protheroe:04} the differential energy spectrum of the accelerated particles ($n=dN/dE$) 
fulfills the equation,
\begin{equation}
\label{Eqn}
\frac{\partial n}{\partial t} + \frac{\partial}{\partial E}(E\ r_{acc}\ n)=Q-r_{esc}\ n,
\end{equation} 
where $r_{acc}$ is the acceleration rate, $r_{esc}$ is the escape rate, and $Q$ is the source term. The solution 
of Eq.~(\ref{Eqn}) for the stationary case ($\partial n_{esc}/\partial t = 0$) and for $Q(E,t)=\dot{N}_0\ \delta(E-E_0)$
is given by,
\begin{equation}
\label{nsol}
n(E) = \frac{\dot{N}_0}{E\ r_{acc}(E)}\ \exp\left( -\int_{E_0}^E dE' \frac{r_{esc}(E')}{E'\ r_{acc}(E')} \right).
\end{equation}
The spectrum of the accelerated particles that escape from the acceleration zone is given by 
$n_{esc}=dN_{esc}/dEdt = r_{esc}\ n$.

If the diffusion coefficients downstream, $k_1$, and upstream, $k_2$, have the same power-law dependence with the energy
of the accelerated particles, $k_1\propto E^{\delta}$ and $k_2\propto E^{\delta}$, the acceleration rate takes the form 
\cite{Protheroe:98},
\begin{equation}
\label{racc}
r_{acc}(E) = a\ E^{-\delta}.
\end{equation}
The escape rate can also be written as \cite{Protheroe:98,Protheroe:04}   
\begin{equation}
\label{resc}
r_{esc}(E) = b\ E^{-\delta} +c\ E^\delta,
\end{equation}
where the first term takes into account the particles that escape downstream and the second term 
is related to the escape of particles due to the finite size of the acceleration region \cite{Protheroe:04}.

From Eqs.~(\ref{nsol}), (\ref{racc}), and (\ref{resc}) the energy spectrum of the escaping particles can be written as,
\begin{eqnarray}
n_{esc}(E) &=& \frac{\dot{N}_0}{E_0} (\gamma-1) \left( 1+\left( \frac{E}{E_{max}}\right)^{2 \delta} \right) %
\exp \left[ -\frac{\gamma-1}{2 \delta} \right. \nonumber \\
&& \times \left. \left(  \left( \frac{E}{E_{max}} \right)^{2 \delta} -\left( \frac{E_0}{E_{max}} %
\right)^{2 \delta}\right) \right],
\label{nesc}
\end{eqnarray}
where $E_{max}=(b/c)^{2 \delta}$ and $\gamma = b/a-1$. Therefore, Eq.~(\ref{Phip}) is obtained from Eq.~(\ref{nesc})
by using
\begin{equation}
\label{Aconst}
A = \frac{\dot{N}_0}{E_0} (\gamma-1) \exp \left[ \frac{\gamma-1}{2 \delta} \left( \frac{E_0}{E_{max}} %
\right)^{2 \delta} \right].
\end{equation}
Note that possible energy losses suffered by the particles during the acceleration process are not taken into 
account in the present calculation.

\end{document}